\documentclass{article}
\usepackage{epsfig}

\newcommand{\beq}{\begin{equation}}
\newcommand{\eeq}{\end{equation}} 
\newcommand{\beqa}{\begin{eqnarray}}
\newcommand{\eeqa}{\end{eqnarray}}
 
\begin{document}

\title{Observation of Bell Inequality violation in B mesons}
\author{Apollo Go \footnote{Corresponding email: apollo.go@cern.ch} \\
\protect\small\em Department of Physics, National Central University 
\\ \protect\small\em Chung-Li, Taiwan \\
Belle Collaboration}

\maketitle
\begin{abstract}
A pair of $B^0\bar B^0$ mesons from $\Upsilon(4S)$ decay exhibit EPR type non-local particle-antiparticle (flavor) correlation. It is possible to write down Bell Inequality (in the CHSH form: $S\le2$) to test the non-locality assumption of EPR. Using semileptonic $B^0$ decays of $\Upsilon(4S)$ at Belle experiment, 
a clear violation of Bell Inequality in particle-antiparticle correlation is observed:
$$ S=2.725\pm0.167_{\rm stat} \pm 0.092_{\rm syst}$$.
\end{abstract}

\section{Introduction}
In Quantum Mechanics (QM), a two-particle non-separable wavefunction lead to a puzzling non-local correlations. This was first pointed out by by Einstein, Podolski and Rosen (EPR) \cite{EPR} in 1935 to challenge the completeness of QM.
A simplified variant was give by David Bohm using spin correlations \cite{EPRB}:
\beq
|\psi\rangle = {1 \over \sqrt{2}} 
        [|\uparrow \rangle_1\otimes  |\downarrow \rangle_2 -
         |\downarrow \rangle_1\otimes  |\uparrow \rangle_2]
\label{singlet}
\eeq
where $|\uparrow \rangle_j$ ($|\downarrow \rangle_j$) describes the spin state of $j^{th}$ particle ($j$=1,2) with spin up (down) respectively. 
Measurement of the state on one particle, 
undetermined prior to the measurement, predicts with certainty the outcome 
of the same state measurement on the second particle. 

In 1964, J.S. Bell put EPR's locality principle into a testable form of an inequality (Bell Inequality) \cite{Bell64}. In 1969, Clauser, Holt, Shimony and Horne \cite{CHSH} translate the Bell Inequality for a pair of entangled photons with two polarization analyzers:
\beq
S=E(\alpha,\beta)-E(\alpha,\beta')+E(\alpha',\beta)+E(\alpha',\beta')\le2
\label{CHSH}
\eeq
where $\alpha$, $\alpha'$ and $\beta$, $\beta'$ are the possible angle settings 
on the polarizers. $E(\alpha,\beta)$ is the correlation function of the polarization measurement outcome on both sides: 
\beq
E(\alpha,\beta)=\int d\lambda\rho(\lambda)\mu(\alpha,\beta,\lambda)=
\int d\lambda\rho(\lambda)\mu(\alpha,\lambda)\mu(\beta,\lambda)
\eeq
where $\mu(\alpha,\lambda)$ is the polarization measurement outcome ($=\pm1$), determined
by the local angle setting $\alpha$ and by a global hidden variable $\lambda$. $\rho(\lambda)$ is a (integrable) probability distribution.
Note the important physical assumption that the
result on one side does not depend on the setting of the other side, 
and vice-versa: $\mu(\alpha,\lambda)$
is independent of $\beta$. This is the locality condition in EPR. 
On the other hand, QM predicts $E(\alpha,\beta)=\cos(\alpha-\beta)$.
In actual experiments, due to the finite efficiency of the detectors, one
does not 
have access to $E(\alpha,\beta)$, only coincidence rates are measurable, such as $R_{++}(\alpha,\beta)$ for the $(+1,+1)$ coincidence rate. Assuming that the
set of detected
events constitute a fair sample, one uses the normalized
correlation function \cite{Aspect80s}
\beq
E_R(\alpha,\beta)=\frac{R_{++}(\alpha,\beta)+R_{--}(\alpha,\beta)-R_{+-}
(\alpha,\beta)-R_{-+}(\alpha,\beta)}
{R_{++}(\alpha,\beta)+R_{--}(\alpha,\beta)+R_{+-}(\alpha,\beta)+R_{-+}
(\alpha,\beta)}.
\label{ER}
\eeq 

For $S$,  it suffice to consider the following
one parameters set ($\theta$) of settings:
$\alpha=0$, $\alpha'=2\theta$, $\beta=\theta$ and $\beta'=3\theta$.
The Bell-CHSH inequality (\ref{CHSH}) then becomes:
\beq
S(\theta)=3E_R(\theta)-E_R(3\theta)\le2
\eeq

Fig.~\ref{fig:CHSH}a  shows the QM prediction of $S(\theta)$, 
there is a clear violation of the Bell-CHSH inequality over the entire 
range $0^o<\theta<68.5^o$ and a maximal violation by a factor $\sqrt{2}$ for
$\theta=45^o$. Many experiments was performed, mostly in testing polarization or spin correlations \cite{Aspect80s,SaclayProton,NISTAtom}. It has never been tested with particle-antiparticle (flavor) correlation.

\section{Bell tests with B mesons}
\begin{figure}
\centerline{\epsfig{figure=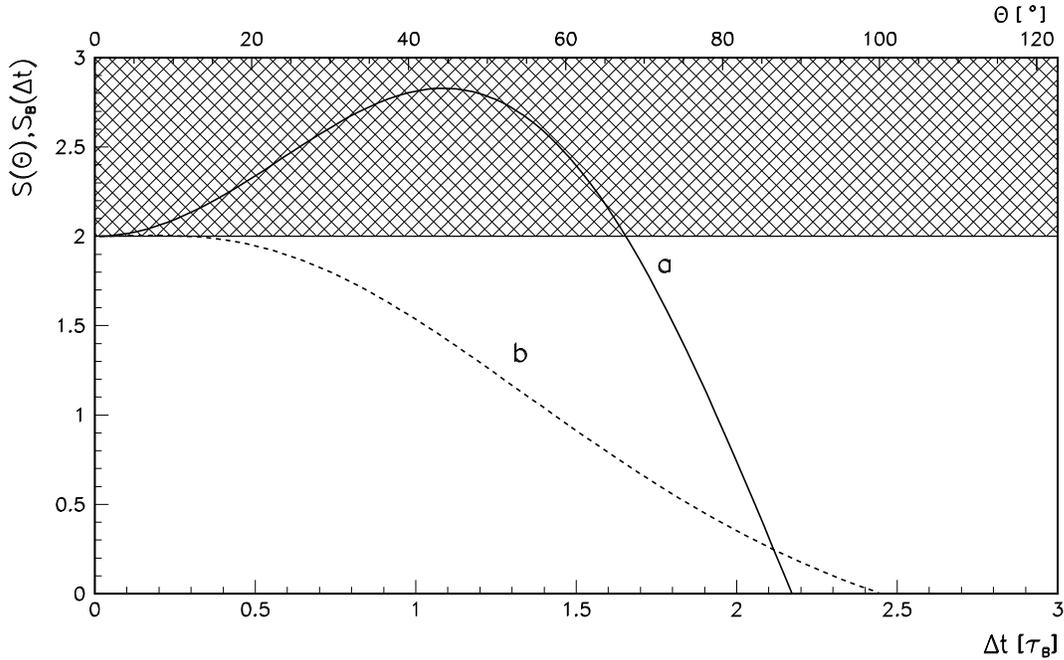,width=6.0in}}
\caption{Bell-CHSH Inequality S for Photon in a singlet state (a) and
        $\Upsilon(4S)\rightarrow B^0\bar{B^0}$ with (b) and without (a) re-normalized decay probability. The shaded area violates Bell Inequality.}
\label{fig:CHSH}
\end{figure}

The wafefunction of $\Upsilon(4S) \rightarrow B^0 \bar B^0$ has exactly the same formalism as the pair of photons (\ref{singlet}):
\beq
|\psi\rangle = {1 \over \sqrt{2}} 
        [|B^0 \rangle_1\otimes  |\bar{B^0} \rangle_2 -
         |\bar{B^0} \rangle_1\otimes  |B^0 \rangle_2]
\label{B0B0b}
\eeq
A measurement of the flavor ($B^0$ or $\bar{B^0}$) on one particle, undetermined prior to the measurement, will determine the flavor of the second particle at the same moment. Comparing to the photons, instead of choosing the polarizer's angle, one measure the flavor at different decay times, $t_a$ and $t_b$, (the effect of mixing is the same as polarizer rotation \cite{GisinGo}). 
The main difference being that $B$ mesons decays which  reduces
coincidence rates \cite{CPLEAR}, e.g.
\beq
R_{B^0B^0}(t_a,t_b)=\frac{1}{4}e^{-2t'/\tau_B} e^{-\Delta t/\tau_B}
\left( 1 - e^{-\Delta t/\tau_B}\cos(\Delta m_d \Delta t)\right).
\eeq
Accordingly the correlations function is:
\beqa
E(t_a,t_b) = - e^{-2t'/\tau_B} e^{-\Delta t/\tau_B} \cos(\Delta m_d \Delta t)
\eeqa
where $\Delta m_d$ is the $B^0-\bar B^0$ mixing parameter, $\tau_B = \tau_{B^0_L}\approx \tau_{B^0_H}=1.542ps$ is the mean $B^0$ decay time, 
$t'={\rm min}(t_a,t_b)$ and $\Delta t=|t_a-t_b|$.
This damping makes it impossible to violate the Bell-CHSH inequality 
(Fig. \ref{fig:CHSH}b) \cite{Ghiraldi}.
However, if one normalizes the correlation function to the undecayed 
pair of $B^0$  (Eq.~\ref{ER}) \cite{Fehrs}, then the correlation 
function is exactly the same as the photon case and depends only on $\Delta t$:
\beqa
E_R(\Delta t) =  \cos(\Delta m_d \Delta t)
\eeqa
The Bell-CHSH Inequality becomes:
\beqa
S(\Delta t)= 3 E_R(\Delta t) - E_R(3\Delta t) \le 2
\eeqa
Below $\Delta t \approx  1.7 \tau _B=2.62ps$, Bell Inequality is violated, just like in photon polarization (Fig. \ref{fig:CHSH}a) \cite{GisinGo}.

\section{Experimental Method}
This measurement is based on data sample of 78 $fb^{-1}$, corresponding to $80 \times 10^6$ $\Upsilon(4S) \rightarrow B\bar B$ decays collected at
Belle detector at the KEKB asymmetric $e^+e^-$ (3.5GeV+8GeV) collider in Japan \cite{KEKB}.  Belle detector consists of a silicon vertex detector (SVD), a central drift chamber (CDC), an array of aerogel threshold Cherenkov counter (ACC), time-of-flight scintillator counters, a CsI(Tl) crystal electromagnetic calorimeter (ECL) located inside a superconducting solenoid magnet of 1.5T. An iron flux-return yoke outside the coil is instrumented to detect $K^0_L$ and to identify muons (KLM) \cite{BELLE}.

We determine the flavor of one neutral $B$ meson by reconstructing
$B^0 \rightarrow D^{*-}l^+\nu $ and $\bar B^0 \rightarrow D^{*+}l^-\nu$ decays where $l^\pm$ denotes lepton (muon or electron). The flavor the other neutral $B$ meson by lepton tagging. Because the $B$ mesons are almost at rest and the boost of the asymmetric collider, $\Delta t$ is proportional to the decay vertexes differences in z (beam) direction: $\Delta t \approx \Delta z/\beta\gamma c$ where $\beta$ and $\gamma$ are the Lorentz boost factor.

\subsection{$B^0 \rightarrow D^{*-}l^+\nu $ Reconstruction}
The decay chains $B^0 \rightarrow D^{*-}l^+\nu, D^{*-} \rightarrow \bar D^0 \pi^-$ with $ \bar D^0 \rightarrow K^+ \pi^-$, $ \bar D^0 \rightarrow K^+ \pi^- \pi^0$ and $ \bar D^0 \rightarrow K^+ \pi^+ \pi^- \pi^0$ (and their charge conjugate modes, same thereafter) were selected.

All charge tracks must have at least one two-dimensional SVD hits and radial impact parameter $|dr| < 0.2$ cm. In addition, we require that $p>0.2 GeV/c$ for tracks in $\bar D^0 \rightarrow K^+ \pi^- \pi^+$ channel. Charged kaons are identified by combining information of the TOF, ACC and $dE/dx$ measurements in the CDC.
$\pi^0$ candidate must have mass within 0.011 $GeV/c^2$ of the $\pi^0$ nominal mass and momentum greater than 0.2 $GeV/c$. The photons from $\pi^0 \rightarrow \gamma \gamma$ decays must have energy greater than $80 MeV/c$.

$\bar D^0$ candidate are selected with $|M_{K\pi(K\pi\pi\pi)}-M_{D^0}|<13MeV/c^2$ for $K^+\pi^-, K^+\pi^-\pi^+\pi^-$ channels. For $K^+ \pi^- \pi^0$ channel,
a cut of $-37MeV/c^2 < (M_{K\pi(K3\pi)}-M_{D^0}) <23MeV/c^2$ where $M_{D^0}=1.865 GeV/c^2$ was applied. Also, the $\pi^0$ with largest Dalitz weight $>10$ was selected to reduce the combinatorial background.

$D^{*-}$ is reconstructed by combining $\bar D^0$ and a low momentum pion. These pion track is re-fitted to the $B^0$ vertex to improve $D^*$ mass resolution. The cut $0.1444 GeV/c^2 < (M_{D^0 \pi_s} - M_{D^0}) < 0.1464 GeV/c^2$ is applied. In addition, center-of-mass (CMS) momentum $P^*_{D^*} > 2.6GeV/c$ is rejected since it is beyond the kinematic limit for $B$ meson decays.

Electrons and muons are with opposite charge to the $D^{*-}$ candidate. Electron identification is based on a combination of CDC $dE/dx$ information, the ACC response, and energy deposit on ECL. Muons are identified by comparing information from the KLM to extrapolated charged particle trajectory. Their momentum are between 1.4 and 2.4 GeV/c in CMS. 

$B^0 \rightarrow D^{*-}l^+\nu $ are selected by combining $D^{*-}$ candidate with muon or electron. The $B^0$ vertex fit must have $\chi^2$ per degree of freedom less than 50 and the fit error less than $100 \mu m$.  The angle in CMS between them greater than 90 degrees. In addition, they must satisfy
$E^*_B-E^*_{D^*l}-|p^*_B|^2-|p^*_{D^*l}|^2+ 2 |p^*_B| |p^*_{D^*l}| \cos(\theta_{B,D^*l})=M_{\nu} \approx 0$ where $\cos(\theta_{B,D^*l})$ is the angle between $p^*_B$ and $p^*_{D^*l}$. A cut $\cos(\theta_{B,D^*l})<1.1$ was imposed.
A cut on $\cos(\theta_{B,D^*l'})>1.1$ was applied where $l'$ is the lepton with momentum reversed. This cut is intended for later for background subtraction \cite{KojiHara}.

\subsection{Flavor tagging}
All other tracks not belonging the above $B^0 \rightarrow D^{*-}l^+\nu $ selection were used for identifying the flavor of the second $B^0$. A multidimensional likelihood method is used and explained in detail in \cite{Btagging}. 
For each flavor decision, we assign a quantity r, which is a MC-determined flavor-tagging dilution factor from 0 to 1. We select events high purity ($r>0.875$). Only lepton tagging sample were used to further increase tagging purity.
The vertex fit must have $\chi^2$ per degree of freedom less than 50 and the fit error less than $140 \mu m$.

\section{Background}
Three types of background have been considered: Continuum, Fake $D^*$ and wrong $D^*$-lepton combination. $4.6 fb^{-1}$ off resonance data were used for continuum background estimation. No event passed the selection. It can be neglected.

\begin{figure}
\centerline{\epsfig{figure=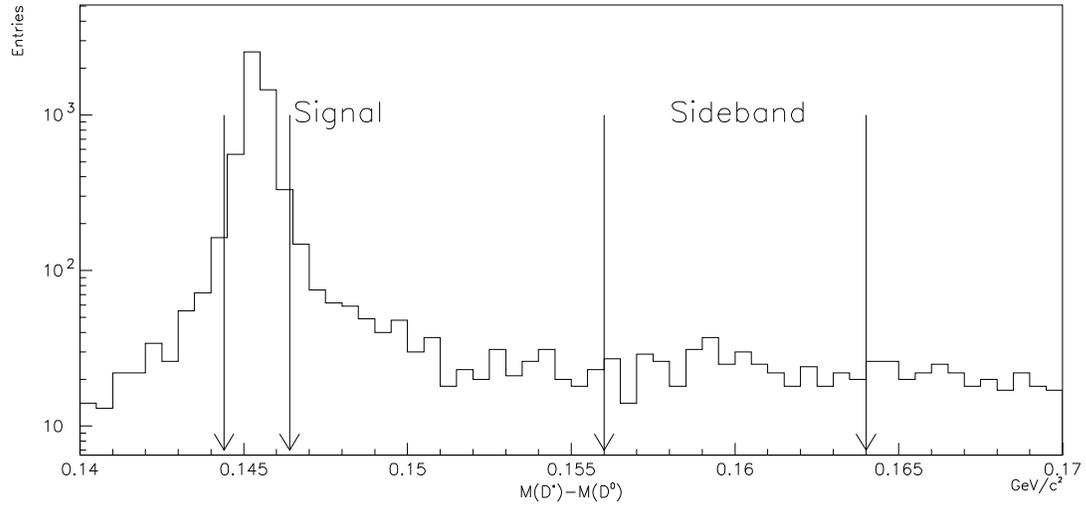,width=6in}}
\caption{Fake $D^*$ background estimation: the signal and sideband region are shown. Notice the logarithmic scale.}
\label{fig:fakeDstar}
\end{figure}

Since $D^{*-}$ is reconstructed from $\bar D^0 \pi^-$ and $ \bar D^0$ from $K^+ \pi^-$, $K^+ \pi^- \pi^0$ and $K^+ \pi^+ \pi^- \pi^0$ decay chains, any fake $D^0$ reconstruction will contribute to the Fake $D^*$ background. We use the sideband of $$0.1560 GeV/c^2 < (M_{D^0 \pi_s} - M_{D^0}) < 0.1640 GeV/c^2$$ to estimate and subtract the number of events under the signal peak (Fig. \ref{fig:fakeDstar}). It is small and has small impact on the value of $S$.

The uncorrelated $D^*-l$ background are mainly due to the reconstruction of a true $D^*$ from one $B^0$ with a lepton from the other $B^0$ with a smaller fraction from Charm decay leptons or fake leptons. We reverse the lepton momentum vector ($l'$) in the CMS and select events where the variable $\cos(\theta_{B,D^*l'})<1.1$. This is to reject correlated $D^*l$ pair while selecting events with no angular correlation and is used in the background subtraction.

\begin{figure}
\centerline{\epsfig{figure=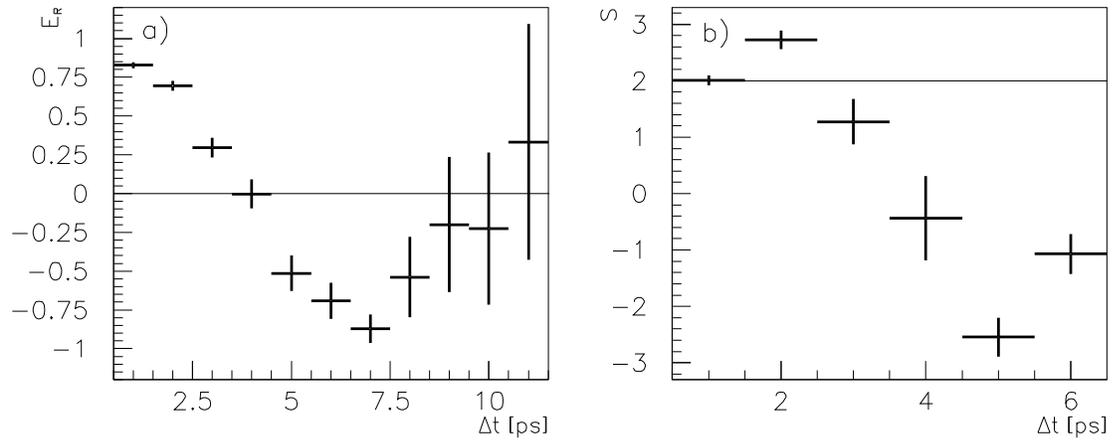,width=6.5in}}   
\caption{Correlation function $E_R$ (a) and Bell Inequality $S$ (b) after event selection and background subtraction.}
\label{fig:AsymS}
\end{figure}

\begin{figure}[htb]
\centerline{\epsfig{figure=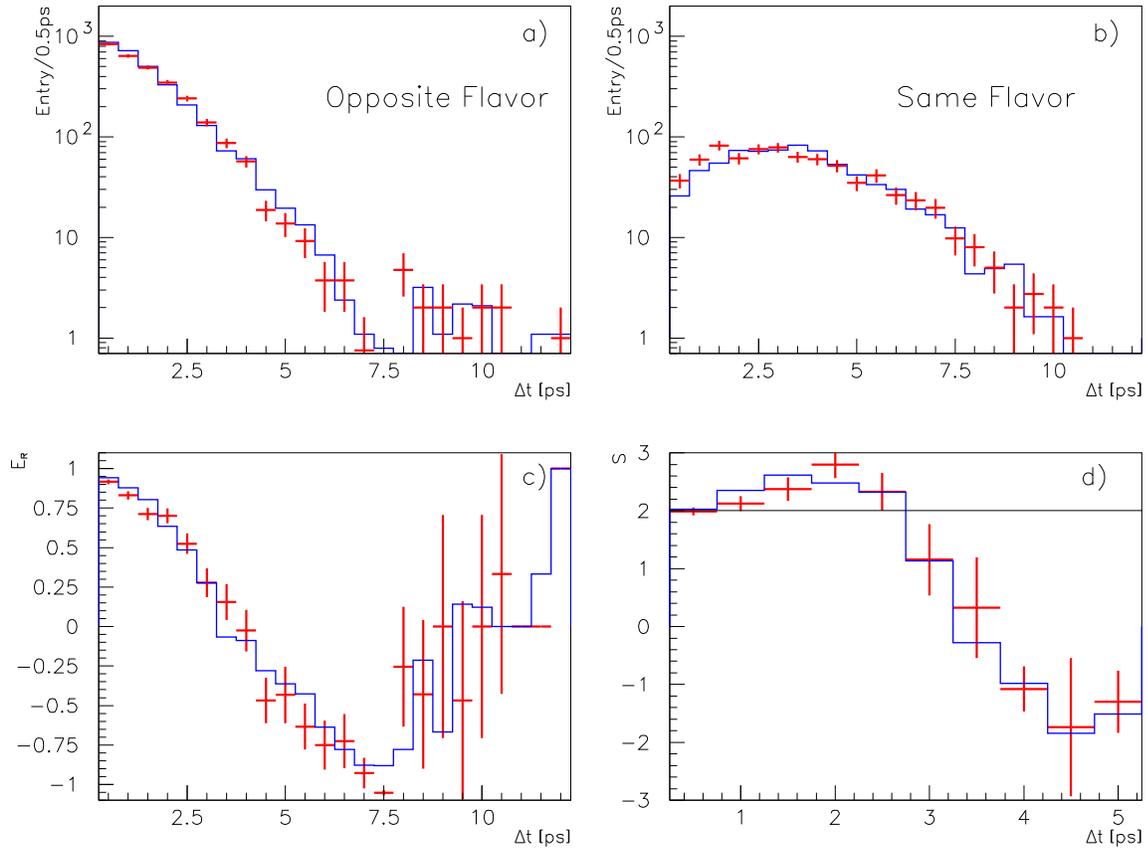,width=6.5in}}   
\caption{Data (crosses) and QM prediction (histogram) for Opposite Flavor (a), Same Flavor (b), $E_R$ (c) and $S$ (d).}
\label{fig:MCdata05ps}
\end{figure}

\bigskip
\noindent
After event selection and backgrounds subtraction, we are left with 3186 events. These events are separated into same flavor (SF) (Fig.~\ref{fig:MCdata05ps}a) and opposite flavor (OF) samples (Fig.~\ref{fig:MCdata05ps}b). Both the correlation function $E_R(\Delta t)$ and $S(\Delta t)$ can be formed (Fig.~\ref{fig:AsymS}). A clear violation of Bell Inequality can be seen  at $\Delta t = (2\pm0.5) ps$:
$$S=2.725 \pm 0.167_{stat},$$

\section{Systematic Error}
Table~\ref{tab:systerr} contains the summary of the estimated systematic error.
The systematic errors for fake $D^*$ and uncorrelated $D^*l$ is estimated from the statistical error, with a $2 \sigma$ variation, of the number of events used in the background subtraction. The rest are estimated by varying the cut parameters by around 20\% (larger than expected variation) from its nominal value.

\bigskip
\noindent
In summary, the final result for S at $\Delta t=(2\pm0.5) ps$ is
$$ S(2\pm0.5ps)=2.725\pm0.167_{\rm stat} \pm 0.092_{\rm syst}$$
which violates Bell Inequality by over $3\sigma$. 

\begin{table}
\center{
\begin{tabular}{lc} \hline \hline
source        & error \\ \hline
Fake $D^*$    & 0.005 \\
uncor. $D^*l$ & 0.030 \\
lepton momentum & 0.060 \\
mistag cut    & 0.030 \\
Particle ID   & 0.028 \\
Vertices Quality & 0.023 \\  
remaining cuts    & 0.042 \\ \hline
total         & 0.092 \\ \hline \hline
\end{tabular}}
\caption{Systematic Error}
\label{tab:systerr}
\end{table}

\section{Comparison with Quantum Mechanics Predictions}
\begin{figure}
\centerline{\epsfig{figure=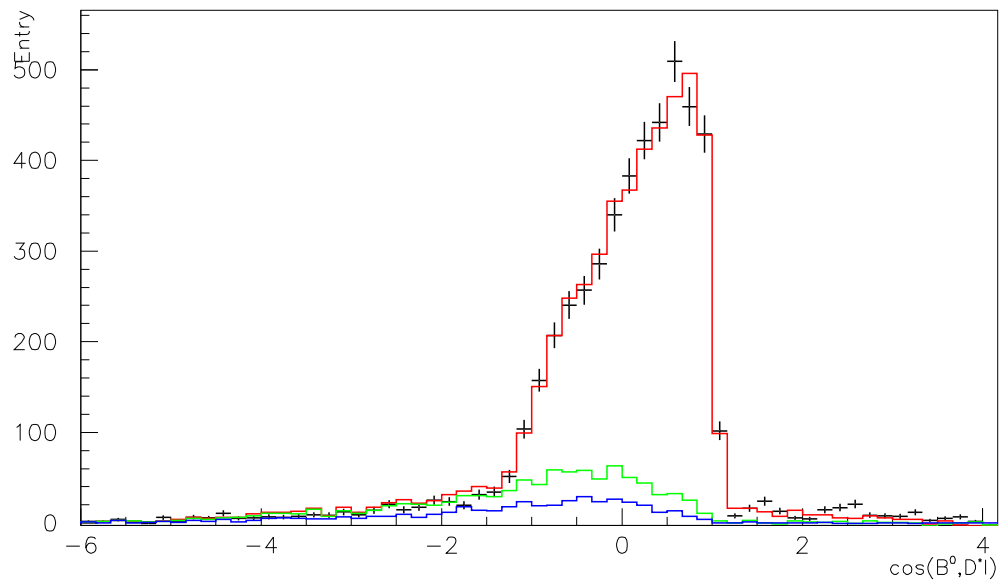,width=6in}}   
\caption{Background level determination from MC $B^\pm(D^{**})$ (bottom histogram), $B^0(D^{**})+B^\pm(D^{**})$ (middle) and total (top) fitted to the data (cross). }
\label{fig:cosfit}
\end{figure}

In addition to the violation of Bell Inequality, we can compare the result with the QM prediction. We use monte-carlo (MC) simulation to generate $B^0 \bar B^0$ pair decays according to QM correlation and folded with detector resolutions and efficiencies. Three types of MC events were generated: $B^0 \rightarrow D^*l\nu$ (signal),
$B^0 \rightarrow D^{**}l\nu$ (it has mixing, considered as signal) and $B^{\pm} \rightarrow D^{0**}l\nu$ (background). We fit these three MC channels to the data (Fig.~\ref{fig:cosfit}) which gives 3.8\% of $B^\pm$ background and 4.5\% of  $B^0 \rightarrow D^{**}l\nu$. Finally, we plot OS, SF, $E_R$ and $S$ for data and for MC (Fig.~\ref{fig:MCdata05ps}). 
Data and MC fits remarkably well, indicating a good agreement between data and QM prediction.

\section{Conclusion}
Bell Inequality is a very powerful test on local hidden variable theories. It has never been tested in particle-antiparticle correlation using massive elementary particles. We measure the Bell-CHSH inequality using semileptonic $B^0$ decays. A violation greater than $3 \sigma$ is observed. MC simulations confirm that the result is consistent with QM predictions. This result will be further improved with the $140 fb^{-1}$ data sample.

\end{document}